\documentclass[conference]{IEEEtran}
\usepackage{graphicx}
\usepackage{url}
\usepackage{amsmath, amsfonts}
\usepackage{amsthm, amssymb}
\usepackage{comment}
\usepackage{xcolor}
\usepackage[font=small]{caption}
\usepackage{subcaption}
\usepackage{array}
\usepackage{dsfont}

\DeclareMathAlphabet\mathbfcal{OMS}{cmsy}{b}{n}

\ifCLASSINFOpdf
\else
\fi
\newcommand\blfootnote[1]{%
  \begingroup
  \renewcommand\thefootnote{}\footnote{#1}%
  \addtocounter{footnote}{-1}%
  \endgroup
}

\begin{document}


\title{Multi-Modal Sensing Aided mmWave Beamforming for V2V Communications with Transformers} 

\author{\IEEEauthorblockN{Muhammad Baqer Mollah$^1$, Honggang Wang$^2$, and Hua Fang$^3$}
\IEEEauthorblockA{$^1$Dept. of Electrical and Computer Engineering, University of Massachusetts Dartmouth, MA 02747 \\ $^2$Dept. of Graduate Computer Science and Engineering, Yeshiva University, NY 10016 \\
$^3$Dept. of Computer and Information Science, University of Massachusetts Dartmouth, MA 02747 \\
Emails: mmollah@umassd.edu, honggang.wang@yu.edu, and hfang2@umassd.edu}
        }

\markboth{}%
{Shell \MakeLowercase{\textit{et al.}}: Bare Demo of IEEEtran.cls for IEEE Journals}

\maketitle

\begin{abstract}
    Beamforming techniques are utilized in millimeter wave (mmWave) communication to address the inherent path loss limitation, thereby establishing and maintaining reliable connections. However, adopting standard defined beamforming approach in highly dynamic vehicular environments often incurs high beam training overheads and reduces the available airtime for communications, which is mainly due to exchanging pilot signals and exhaustive beam measurements. To this end, we present a multi-modal sensing and fusion learning framework as a potential alternative solution to reduce such overheads. In this framework, we first extract the features individually from the visual and GPS coordinates sensing modalities by modality specific encoders, and subsequently fuse the multimodal features to obtain predicted top-$k$ beams so that the best line-of-sight links can be proactively established. To show the generalizability of the proposed framework, we perform a comprehensive experiment in four different vehicle-to-vehicle (V2V) scenarios from real-world multi-modal sensing and communication dataset. From the experiment, we observe that the proposed framework achieves up to $77.58$\% accuracy on predicting top-$15$ beams correctly, outperforms single modalities, incurs roughly as low as $2.32$~dB average power loss, and considerably reduces the beam searching space overheads by $76.56$\% for top-$15$ beams with respect to standard defined approach. \blfootnote{This paper has been accepted to present at 2025 IEEE Global Communications Conference (GLOBECOM), Taipei, Taiwan.}
\end{abstract}


\IEEEpeerreviewmaketitle

\section{Introduction}
    \blfootnote{\copyright 2025 IEEE. Personal use of this material is permitted. Permission from IEEE must be obtained for all other uses, in any current or future media, including reprinting/republishing this material for advertising or promotional purposes, creating new collective works, for resale or redistribution to servers or lists, or reuse of any copyrighted component of this work in other works.} Future transportation systems will be revolutionized by connected and autonomous vehicles through greatly improving traffic efficiency, mobility, and transportation safety, where seamless connectivity with high throughput and lower latency is crucial for extensive sensor data collection and sharing among other vehicles, infrastructures, and surroundings \cite{cheng2025driving}. Such sensor data sharing play an important role in safety critical tasks to support self-driving functionalities cooperative perception for managing blind spots. For example, from report \cite{misc1}, an approximate $25$ gigabytes/hour is required by a connected vehicle for the purpose of sensor data communication.

    
    While the the 5G New Radio (5G-NR) standard has specified to operate in both sub-6 GHz and higher frequencies ($> 24$ GHz) at millimeter-wave (mmWave) band, the sub-6 GHz band is already too congested to use in these instances. Rather, the mmWave band has been considered as an ideal candidate due to having an abundant spectrum resources toward fulfilling the high throughput and lower latency demands \cite{schott2024mm,mollah2024mmwave}. In general, signals in mmWave band suffer from severe atmospheric path loss due to having short wavelengths (roughly 1$\sim$10 mm) \cite{osa2023measurement}. For this reason, beamforming techniques are utilized in mmWave systems by deploying phased-antenna arrays and configuring a narrow beam of RF energy to establish and maintain high directional links, thereby ensuring sufficient signal strength.
    
    On the other hand, the 5G-NR standard defines a beam training approach based on codebooks over phased-antenna arrays \cite{xue2024survey}. Typically, the codebook-based approach enables the communicating nodes to coordinate each other by exchanging pilot signals and measurements for selecting the best beam direction. However, when it comes to vehicular moving scenarios where the beam directions change dynamically, this process often introduces highly undesirable beam training overheads. Beside, the overheads may also increase linearly with the number of beam directions in codebooks. Accordingly, it becomes essential to reduce such beam training overheads in order to fully unlock the potential benefits from mmWave communications in connected vehicles domain.
    
    Considering the aforementioned challenges, recent research has suggested utilizing out-of-band contextual information as a promising alternative to perform beam selection, where the contextual information can obtained from non-RF sensing, such as GPS \cite{mollah2024position}, visual \cite{wang2025deep}, or LiDAR \cite{ohta2025real2sim2real} sensors. As an example, the work in \cite{salehi2022flash} has showed the average end-to-end latency on beam selection can be possibly reduced by 52.75\% on average time by utilizing such vehicular sensing information, while the work \cite{salehi2022deep} has observed that similar sensing information can help to reduce the beam searching overheads by 80\% for 5G-NR.
    
    Over the recent works, several approaches have been introduced on leveraging multi-modal sensing to support mmWave beamforming. For example, the authors in \cite{charan2025camera} have introduced multi-modal machine learning model on visual and GPS data to perform beam prediction in multi-candidate wireless environments. Beside, the works in \cite{zhang2024integrated} have focused on deep learning based multimodal fusion of sensing data for proactive beamforming. In particular, this work utilized two regression models to handle the high-dimensional inputs and account for random vehicle drifting to predict the future position of targeted vehicle, where each trajectory point has visual sensing information. Meanwhile, Kim et al. \cite{kim2024vision} have proposed a strategy, where the authors have focused on minimizing the beam training overheads by estimating the user's position accurately by wireless sensing and visual data. In another work \cite{raha2025advancing}, the authors have utilized visual and location information, however, focused more on enhance the robustness of the beamforming approach considering visual data in varying weather conditions, where a transformer network and semantic localization techniques are incorporated to do so.
    
    However, the aforementioned contributions are largely limited to either by ray-tracing simulations, or developed for vehicle-to-infrastructure (V2I) communications, without exploring on real-world measured sensing data and the vehicle-to-vehicle (V2V) communications like highly dynamic scenarios. To this end, this motivates us to propose a transformer based multi-modal solution for mmWave enabled V2V communications. In particular, the proposed framework utilizes targeted vehicle's position along with captured visual sensing information to estimate a limited set of candidate beams, that is top-$k$ beams, subsequently reducing the beam search space. Most importantly, the proposed solution can fully leverage the sensors installed on the vehicles, thereby simplifying the complicacy. Finally, we validate the proposed framework with on 60 GHz mmWave sensing datasets in four different scenarios, and present the findings in terms of prediction accuracies and average power losses matrices. The performance evaluation results show that utilizing multimodal fusion approach outperforms than the individual modalities, which is due to capturing the features more comprehensively by using more than one modality.
    
\section{System Model and Problem Description}
    In this section, we first present the considered system model, followed by formally describing the beam selection problem.
    
\subsection{System Model}
\begin{figure} [!t]
    \includegraphics[width=\linewidth]{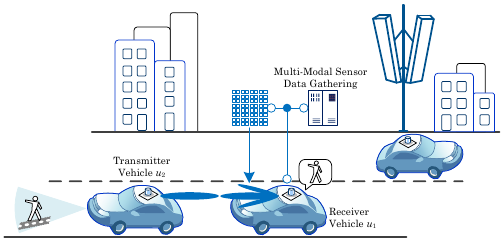}
    \centering
    \caption{Illustration of our considered system model enabled by mmWave V2V communications.}
    \label{fig: system}
\end{figure}

    In this work, we consider a downlink vehicular communication system with two moving vehicles $u_1$ and $u_2$, respectively. As illustrated in Fig.~\ref{fig: system}, the moving vehicle $u_1$ is connected by vehicle-to-vehicle (V2V) communication link to the another vehicle $u_2$ to enable communication services, such as cooperative perceptions withing the coverage area. Specifically, the unit $u_1$ is acting as receiver, while the unit $u_2$ is acting as transmitter. In order to capture multi-modal sensing information from the surrounding environment, the receiver units are equipped with camera sensor, whereas the transmitter unit carry the GPS receiver.
    
    Here, the $u_1$ is assumed to be equipped with (an) phased arrays employing $N_{rx}$ antenna elements uniform linear arrays (ULA) operating at mmWave frequency bands. Similarly, the other unit $u_2$ is equipped with $N_{tx}$ elements antennas. For the purpose of simplicity, $N_{tx} \gg N_{rx}$ is assumed, and subsequently we set $N_{tx} = 1$, which refers the omnidirectional transmissions are realized by using the single antenna element of the phased array. However, each phased arrays at receiver sides employs a pre-defined over-sampled beamsteering codebooks $\mathbfcal{C}_{rx} = \{\mathsf{c}_1, \mathsf{c}_2, ...., \mathsf{c}_K\}$ containing $K$ number of total beam directions toward receiving the transmitted signals. Here, $\mathsf{c}_{\kappa} \in \mathbb{C}^{N_{rx} \times 1}$ represents any beamsteering vector.
    
    Further, we consider the downlink communication systems adopting orthogonal frequency-division multiplexing signal transmissions model. Let, $\mathcal{Q} = \{1, ...., q_m\}$ represents the set of subcarrier indices with $q_m$ total number of subcarriers in a time-varying channel, $s \in \mathbb{C}$ is the transmitted data symbol, $\mathbb{E}|s|^2$ = $P_s$ (average power), and $\mathbf{h} \in \mathbb{C}^{N_{rx} \times 1}$ indicates the downlink channels of the V2V link. 
    
\subsection{Problem Description}
    Given the considered system model, the receiving vehicle $u_1$ observes the corresponding downlink received signal at time $t$ and $q^{\text{th}}$ subcarrier as $r_t[q] = \mathbf{h}_t^\mathsf{T}[q] \mathsf{c}_{\kappa} s_t[q] + n_t[q], \quad \forall q \in \mathcal{Q},$
    where, $(\cdot)^\mathsf{T}$ is the conjugate transpose, and the variable $n \sim \mathcal{C}\mathcal{N}(0, \sigma_n^{2})$ denotes as the received complex Gaussian noise sample with zero mean and $\sigma_n^{2}$ variance. With the fixed codebook constraint, let $\mathcal{B}_{rx} = \{\mathsf{b}_1, \mathsf{b}_2, ...., \mathsf{b}_K\}$ with $|\mathcal{B}_{rx}| = K$ is the set of possible all beams to perform beamforming by the vehicle $u_1$ to obtain adequate received powers, and $\mathcal{P}_{rx} = \{\mathsf{p}_1, \mathsf{p}_2, ...., \mathsf{p}_K\}$ is the set of corresponding normalized received power, where the received power for a specific beam $\mathbf{p}_{\kappa}$ can be calculated by summing over all subcarriers. Then, the beam selection task is to find the optimal beam $\mathsf{b}_{\kappa}^*$ out of $\mathcal{B}_{rx}$ that maximizes the received power. Mathematically, this beam selection problem can be formulated as follows
\begin{equation}
    \mathsf{b}_{\kappa}^* = \underset{\mathsf{c}_{\kappa} \in \mathcal{C}_{rx}}{\arg\max} \sum_{q=0}^{\mathcal{\mathcal{Q}}-1} |\mathbf{h}_t^\mathsf{T}[q] \mathbf{c}_{\kappa} s_t[q]|^2.
\end{equation}

    In 5G-NR standard defined beam selection approach, finding the optimal beam $\mathsf{b}_{\kappa}^*$ requires sweeping all beams sequentially over the beam codebook, which typically introduces high beam training and latency overheads. However, the aim of this work is to investigate how to utilize out of band sensing information to address this problem by predicting a set of recommended beams $\mathcal{B}_{k} = \{\mathsf{b}_i\}_{i=1}^k$ as top-$k$ beams such that $\mathcal{B}_{k} \subset \mathcal{B}_{rx}$ and $\mathsf{b}_{\kappa}^* \in \mathcal{B}_{k}$ as top-1 beam.

\begin{figure*} [!t]
	\includegraphics[width=.85\linewidth]{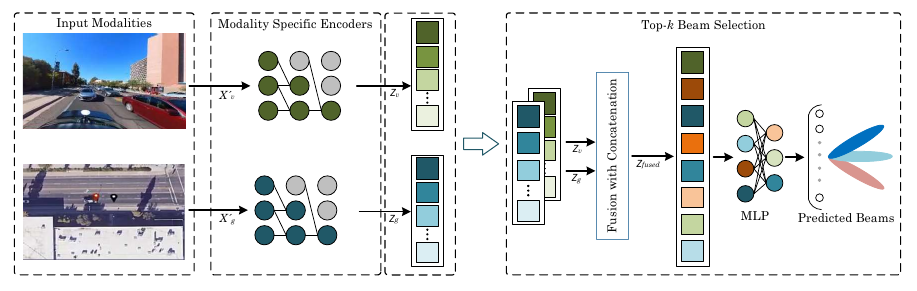}
    \centering
    \caption{The overview of the proposed transformer based framework for top-$k$ beam predictions.}
    \label{fig: transformer-Model}
\end{figure*}

\section{The Proposed Framework}
    In this section, we describe the design overview and key steps of our proposed transformer based beam selection framework as presented in Fig. \ref{fig: transformer-Model}, which is comprised of two distinct modality specific encoders for feature extractions, each taking the pre-processed modalities as inputs, followed by multi-modal fusion and beam selection procedures.

\subsection{Solution Methodology}
    The proposed framework leverages multimodal sensing modalities, which are captured from the following three sensing elements, such as position sensors, visual sensors, and wireless sensors. We define the multi-modal samples captured by the aforementioned sensors at timestamp $t$ as $X_{g}[t] \in \mathbb{R}^{2}$ (latitude and longitude), $X_{v}[t] \in \mathbb{R}^{w_v \times h_v \times c_v}$ (width, height, and number of color channels), and $Y_i$, respectively, where $Y_i$ in particular represents the optimal beam label across the codebook. During the data acquisition phase, the connected vehicle $u_1$ are required to collect those samples in different scenarios and subsequently build a dataset, $\mathcal{D} = \{(X_i, Y_i)\}_{i=1}^N$, where $X_i = (X_{g}, X_{v})$ and $N$ is the number of total samples.
    
    Once the dataset $\mathcal{D} = \{(X_i, Y_i)\}_{i=1}^N$ has been built, we solve the beam selection task as defined in Section II-B by a proposed multi-modal sensing assisted framework. For that, we set up a multimodal model $\mathcal{M}(\cdot, \omega)$ to learn the mapping of the multi-modal sensing inputs (position and visual) to the top-$k$ beams, where $\omega$ is the learnable parameters which can be learned (i.e., adjusted) by training. Toward this objective, the multimodal model can be optimized to find the optimal model parameter $\omega^*$ with the aim of minimizing the loss between the model output and optimal beam labels (ground truth beams). Mathematically, the optimization problem can be expressed as follows:
\begin{equation}
    \omega^* = \underset{\omega}{\arg\min} \frac{1}{|\mathcal{D}|}\sum_{(X_i, Y_i) \in \mathcal{D}} \mathcal{J}(\mathcal{M}(X_i, \omega), Y_i),
\end{equation}
    where, $\mathcal{J}(\cdot)$ represents the loss function which measures the difference between predicted (estimated) beams from the model and ground truth beams. However, toward solving the problem defined in Eq. (3), we utilize cross entropy as the loss function during the training phase given by $\mathcal{J}(\hat{\mathsf{b}}, Y) = - \frac{1}{N}\sum_{i=0}^{N-1}\sum_{k=0}^{K-1} Y_{ik}\log(\hat{\mathsf{b}}_{ik}),$
   where $K$ and $\hat{\mathsf{b}}$ represent the total number of beams and the predicted output from the model, respectively. After the model training process, the most up-to-date trained model can be utilized to infer beams by the vehicle $u_1$ to serve their respective connected vehicle $u_2$ within their coverage areas.
   
\subsection{Modality Specific Encoders}    
    The collected raw modality samples are desired to be compatible to the encoder and properly formatted while being used as inputs to the encoders. Hence, we first pass through the modalities into following preprocessing steps. 
    
    For position sample preprocessing, we apply a normalization technique across the latitude and longitude values with a min-max scaling. And for the visual sample, we first resize the sample images into a reduced and same sized spatial dimensions of $224 \times 224$, where $224 \leq w_v$ and $224 \leq h_v$, then the total numbers of pixels become $224 \times 224 = 49,536$. After that, we normalize the red, green, and blue color channel intensities (each range from 0-255) to have zero mean and unit variance by subtracting a fixed mean $(0.485, 0.456, 0.406)$ and dividing the results by a fixed standard deviation $(0.229, 0.224, 0.225)$ for each color channels. To this end, we can express the aforementioned preprocessed multi-modal samples at timestamp $t$ as $X^{\prime}_{g}[t]$ and $X^{\prime}_{v}[t]$, respectively. After preprocessing steps, we then extract the features independently by employing the following modality specific encoders with focusing on extracting the most relevant features from each modality.

    \textit{1) Position Encoder:} For constructing the position encoder, we adopt two components from the transformer architecture: an embedding layer and a transformer encoder, and customize accordingly for GPS coordinates feature extraction modeling. To be specific, given the preprocessed position data $X^{\prime}_{g} \in \mathbb{R}^2$, the embedding layer first transforms each input feature from $X^{\prime}_{g}$ into a higher-dimensional feature space (i.e., $d_g = 512$) in order to enable richer feature learning, which can be expressed as $\mathbf{E}_g = X^{\prime}_{g}\mathbf{W}_{g} + \mathbf{b}_{g},$ where $\mathbf{W}_g \in \mathbb{R}^{d_g \times 2}$ and $\mathbf{b}_{g}$ are the weight matrix and bias, respectively. After obtaining the higher-dimensional features, we then utilize a multi-layered transformer encoder equipped with $6$ stacked layers, each layer composed of a multi-head attention sub-layer (each with $h$ = $8$ number of attention heads), followed by a feed-forward sub-layer. In particular, for the calculation of attention score for single attention, the respective query ($\mathbf{Q}_g$), key ($\mathbf{K}_g$), and value ($\mathbf{V}_g$) matrices are derived from the output of embedding layer as $\mathbf{Q}_g = \mathbf{E}_g\mathbf{W}_{g}^{Q}, \quad \mathbf{K}_g = \mathbf{E}_g\mathbf{W}_{g}^{K}, \quad \mathbf{V}_g = \mathbf{E}_g\mathbf{W}_{g}^{V}.$ 
    Here, $\mathbf{W}_{g}^{Q}$, $\mathbf{W}_{g}^{K}$, and $\mathbf{W}_{g}^{V}$ are the learnable parameters during training, whereas the dimensionality of both $\mathbf{Q}_g$ and $\mathbf{K}_g$ is $d$ and $\mathbf{V}_g$ is $d_v$. Subsequently, we can get the attention score by computing the following scaled dot-product self-attention operation equation:
\begin{equation}
    \mathsf{Attention}(\mathbf{Q}_g, \mathbf{K}_g, \mathbf{V}_g) = \mathsf{Softmax}\left(\frac{\mathbf{Q}_g\mathbf{K}_g^{\mathsf{T}}}{\sqrt{d}} \right)\mathbf{V}_g.
\end{equation}
    
    However, aiming to capture richer contextual relationships by focusing on the different parts of the input at once, the attention operation (or head) are calculated $h$ times in parallel within the multi-head attention. This can be obtained by linearly transforming (also referred as projection) the $\mathbf{Q}_g$, $\mathbf{K}_g$, and $\mathbf{V}_g$ with $h$ times by using different learnable matrices. After that, considering the dimensions $d = d_v = d_g/h = 64$ for each of these, the results from all heads are concatenated and projected once again as $\mathsf{MHA}(\mathbf{Q}_g, \mathbf{K}_g, \mathbf{V}_g) = \mathsf{Concat}(head_1,...,head_h)\mathbf{W}^o,$
    where $head_l = \mathsf{Attention}(\mathbf{Q}_{gl}, \mathbf{K}_{gl}, \mathbf{V}_{gl})$ is the $l$-th head and $\mathbf{W}_o \in \mathbb{R}^{hd_v \times d_g}$ is the parameter matrix used for the projection. Finally, we pass the multi-head attention results through the feedforward network sub-layer to improve the feature expressiveness, thereby producing the desired features, as $\mathsf{Z}_{g}$.
    
    \textit{2) Visual Encoder:} To implement the visual encoder, we employ a vision transformer variant, named multi-axis vision transformer (MaxViT) model \cite{tu2022maxvit} and fine-tune according to our task. This model introduces a unified design by taking benefits from both convolution and transformer to extract rich visual features (representations) while better adapting capabilities to high-resolution and dense prediction tasks. Particularly, the MaxViT model comprises of three components, such as convolution, mobile inverted bottleneck convolution (MBConv), and multi-axis attentions. With these components, the model takes $X^{\prime}_{v}$ as input and processes it through the following ways.
    
    At first stage, the convolutional operations with $3 \times 3$ kernels are performed on the input $X^{\prime}_{vis}$ for initial feature extractions to capture simple basic features, and also downsampled with a stride of $2$. The resulting outputs are then passed through a series of MaxViT blocks across four stages, where each MaxViT block incorporates the MBConv and multi-axis attentions. Here, the MBConv uses depthwise separable convolution with an additional squeeze and excitation, and follows inverter residual block structure. On the other hand, the multi-axis attentions in MaxViT blocks basically consist of two types of attention mechanisms, such as block attention and grid attention to obtain local and global interactions, respectively. Let $\digamma_v \in \mathbb{R}^{w_{v}^{\prime} \times h_{v}^{\prime} \times c_{v}^{\prime}}$ is the intermediate feature map toward the attentions, and the shape of the blocks after partitioning the features into non-overlapping windows becomes $(w_{v}^{\prime}/\mathbf{w} \times h_{v}^{\prime}/\mathbf{w}, \mathbf{w} \times \mathbf{w}, \times c_{v}^{\prime})$, denoting $\digamma_{v\mathbf{w}}$ and each of ($\mathbf{w} \times \mathbf{w}$) size. Then, applying attention within each window independently, that is, the local spatial dimension ($\mathbf{w} \times \mathbf{w}$), we can express as follows.
\begin{equation}
    \mathsf{Attention}(\mathbf{Q}_{v\mathbf{w}}, \mathbf{K}_{v\mathbf{w}}, \mathbf{V}_{v\mathbf{w}}) = \mathsf{Softmax}\left(\frac{\mathbf{Q}_{v\mathbf{w}}\mathbf{K}_{v\mathbf{w}}^{\mathsf{T}}}{\sqrt{d}} \right)\mathbf{V}_{v\mathbf{w}},
\end{equation}
    where, $\mathbf{Q}_{v\mathbf{w}} = \digamma_{v\mathbf{w}}\mathbf{W}_{v\mathbf{w}}^{Q}$, $\mathbf{K}_{v\mathbf{w}} = \digamma_{v\mathbf{w}}\mathbf{W}_{v\mathbf{w}}^{K}$, and $\mathbf{V}_{v\mathbf{w}} = \digamma_{v\mathbf{w}}\mathbf{W}_{v\mathbf{w}}^{V}$. And, we can get the multi-head attention within small local windows as: $\mathsf{Z}_{local} = \mathsf{MHA}(\mathbf{Q}_{v\mathbf{w}}, \mathbf{K}_{v\mathbf{w}}, \mathbf{V}_{v\mathbf{w}})$. 
    
    Following the block attention, we assume the shape of the grids $\digamma_{v\mathbf{g}}$ is $(\mathbf{g} \times \mathbf{g}, w_{v}^{\prime}/\mathbf{g} \times h_{v}^{\prime}/\mathbf{g}, \times c_{v}^{\prime})$. Then, the grid attention is applied on a sparsely sampled uniform grid ($\mathbf{g} \times \mathbf{g}$), which can be calculated as:
\begin{equation}
    \mathsf{Attention}(\mathbf{Q}_{v\mathbf{g}}, \mathbf{K}_{v\mathbf{g}}, \mathbf{V}_{v\mathbf{g}}) = \mathsf{Softmax}\left(\frac{\mathbf{Q}_{v\mathbf{g}}\mathbf{K}_{v\mathbf{g}}^{\mathsf{T}}}{\sqrt{d}} \right)\mathbf{V}_{v\mathbf{g}},
\end{equation}
    where, $\mathbf{Q}_{v\mathbf{g}} = \digamma_{v\mathbf{g}}\mathbf{W}_{v\mathbf{g}}^{Q}$, $\mathbf{K}_{v\mathbf{g}} = \digamma_{v\mathbf{g}}\mathbf{W}_{v\mathbf{g}}^{K}$, and $\mathbf{V}_{v\mathbf{g}} = \digamma_{v\mathbf{g}}\mathbf{W}_{v\mathbf{g}}^{V}$. After this, we can obtain the multi-head attention across non-overlapping grid stripes for long-range interactions across the entire image by: $\mathsf{Z}_{global} = \mathsf{MHA}(\mathbf{Q}_{v\mathbf{g}}, \mathbf{K}_{v\mathbf{g}}, \mathbf{V}_{v\mathbf{g}})$. It should be noted that the same fixed sizes ($\mathbf{w} = \mathbf{g} = 7$) are considered for both window and grid attentions, while the attention head size is set to be $32$, and a standard feed-forward network in applied after each attentions. Based on this MaxViT block including MBConv, block attention, and grid attention, the overall MaxViT model builds a typical hierarchical structure by simply stacking four identical MaxViT blocks, however, each block has half resolution of the previous block along with a doubled channels.

\begin{figure*} [!t]
\centering
\begin{subfigure}[b]{0.24\textwidth}
    \centering
    \includegraphics[width=4.4cm]{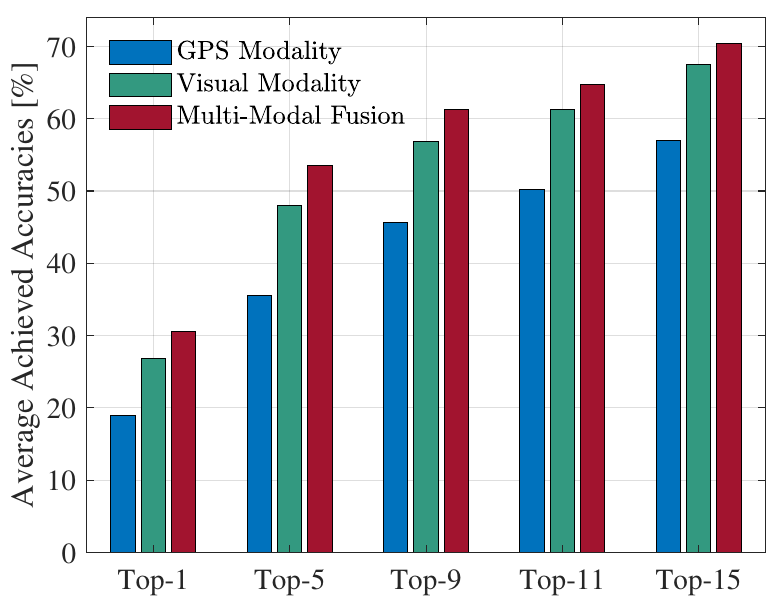}
	\caption{\centering \scriptsize For V2V-Day Scenario}
\end{subfigure}
\begin{subfigure}[b]{0.24\textwidth}
	\centering
	\includegraphics[width=4.4cm]{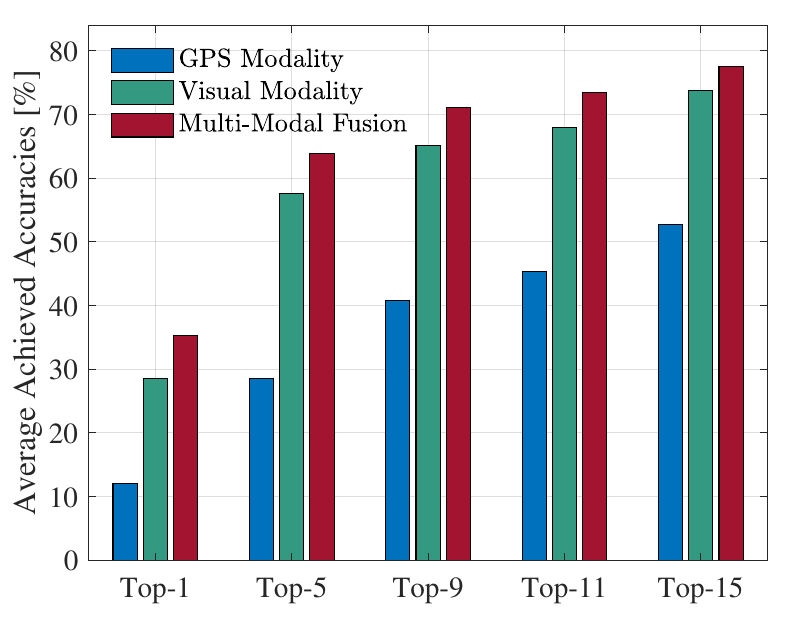}
	\caption{\centering \scriptsize For V2V-Night Scenario}
\end{subfigure}
\begin{subfigure}[b]{0.24\textwidth}
    \centering
    \includegraphics[width=4.4cm]{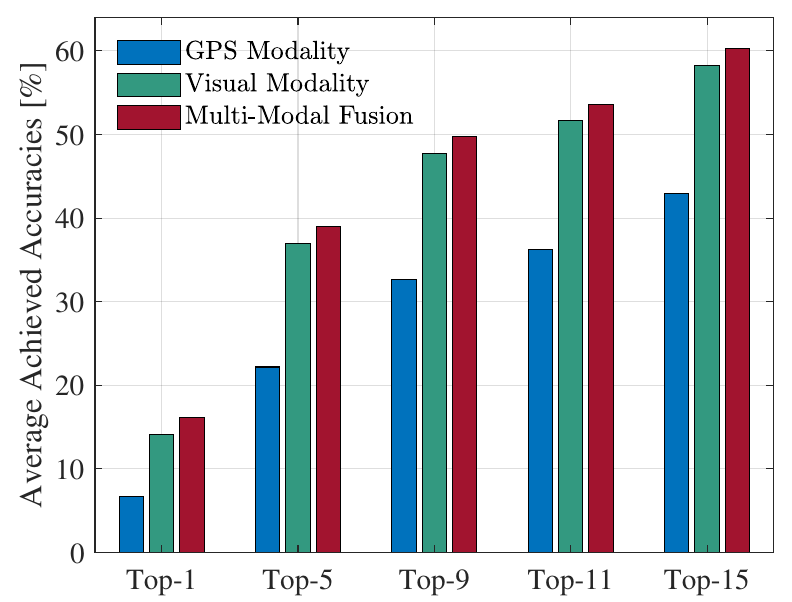}
    \caption{\centering \scriptsize For V2V-Day Scenario}
\end{subfigure}
\begin{subfigure}[b]{0.24\textwidth}
    \centering
    \includegraphics[width=4.4cm]{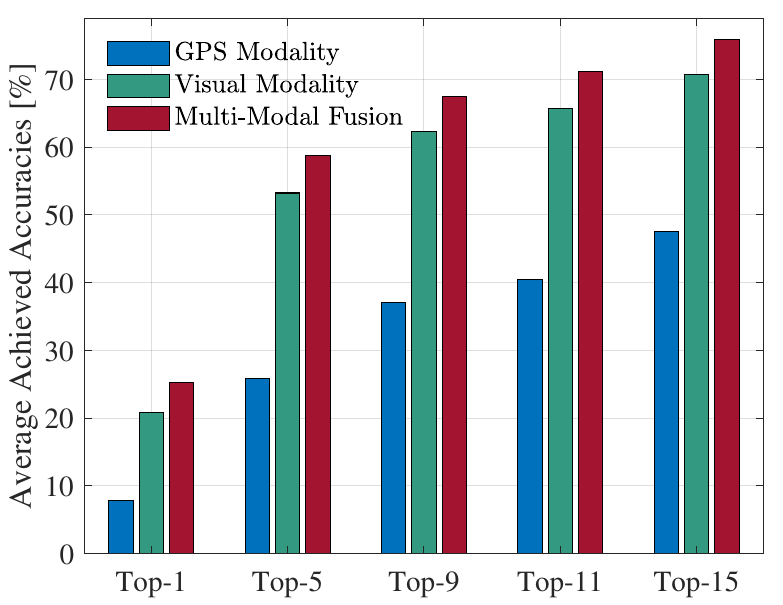}
    \caption{\centering \scriptsize For V2V-Night Scenario}
\end{subfigure}

\vspace{3mm}

\begin{subfigure}[b]{0.24\textwidth}
    \centering
    \includegraphics[width=4.4cm]{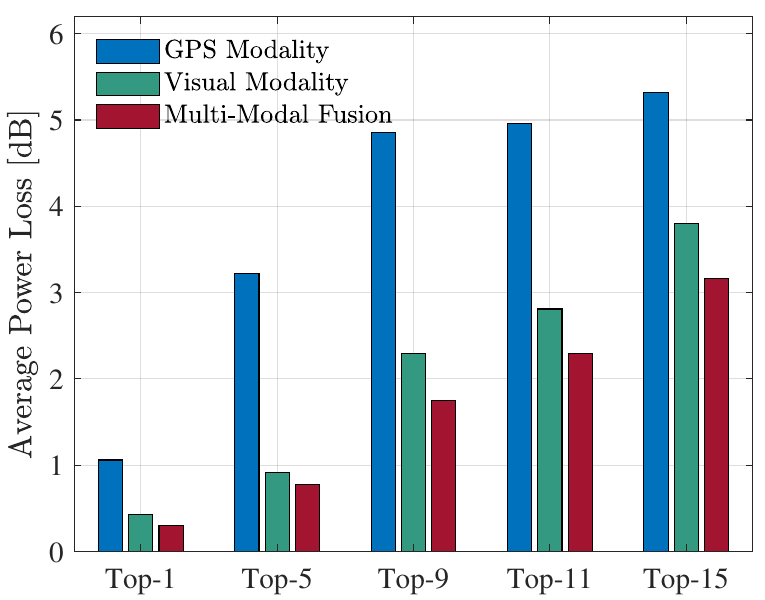}
	\caption{\centering \scriptsize For V2V-Day Scenario}
\end{subfigure}
\begin{subfigure}[b]{0.24\textwidth}
	\centering
	\includegraphics[width=4.4cm]{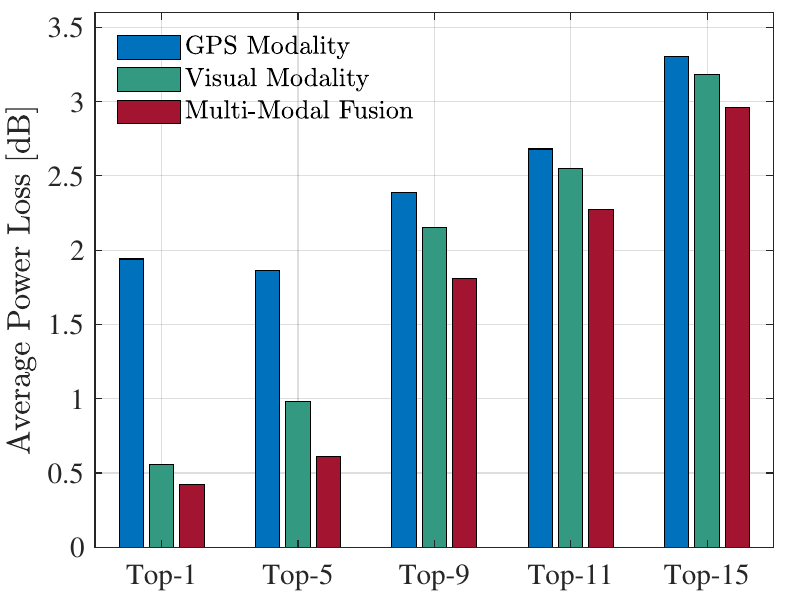}
	\caption{\centering \scriptsize For V2V-Night Scenario}
\end{subfigure}
\begin{subfigure}[b]{0.24\textwidth}
    \centering 
    \includegraphics[width=4.4cm]{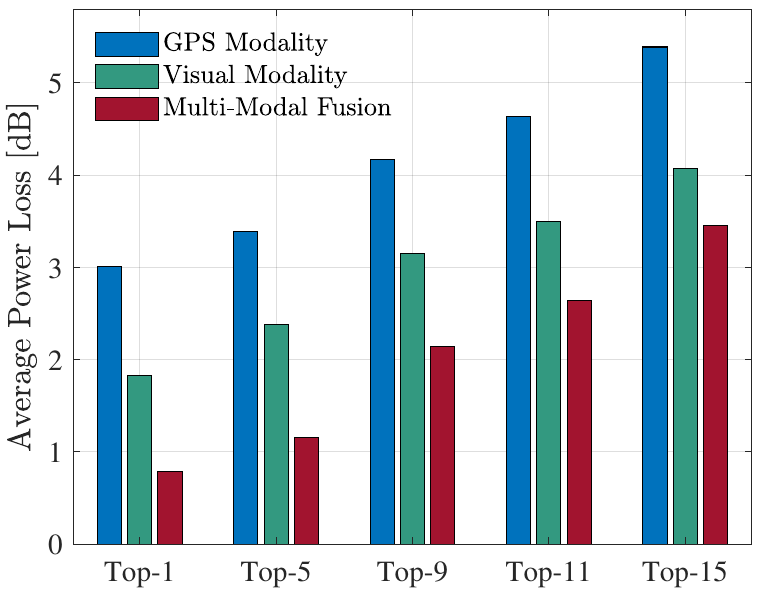}
    \caption{\centering \scriptsize For V2V-Day Scenario}
\end{subfigure}
\begin{subfigure}[b]{0.24\textwidth}
    \centering
    \includegraphics[width=4.4cm]{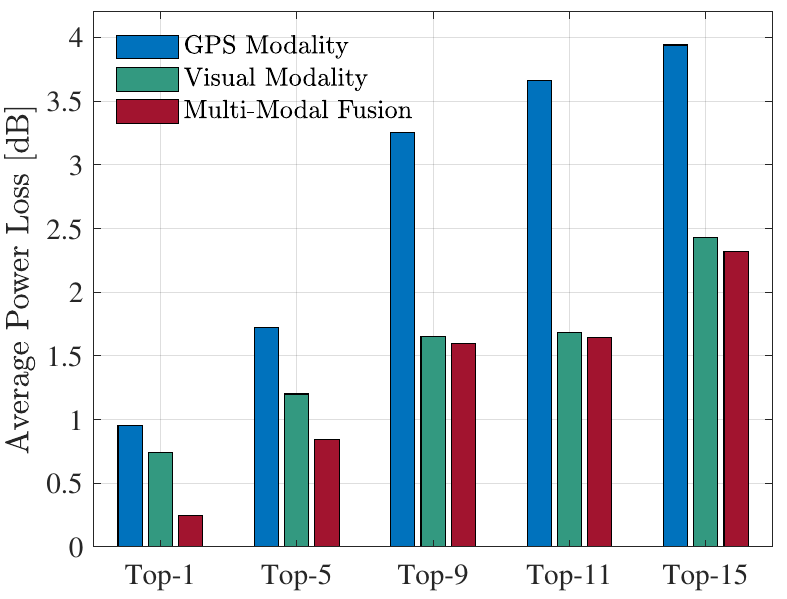}
    \caption{\centering \scriptsize For V2V-Night Scenario}
\end{subfigure}
    \caption{The performance comparisons on average achieved accuracies in percentages ($\%$) and average power mmWave loss in decibels (dB) tested on all considered V2V scenarios.}
	\label{fig: Performances}
\end{figure*}

\subsection{Top-$k$ Beam Selection}
    Let the respective outputs from each encoders as $\mathsf{Z}_g$ and $\mathsf{Z}_v$, the beam selection module produces the final prediction results. For that, we first carry out the feature fusion along their feature dimensions through a concatenation function to obtain a unified representation as $\mathsf{Z}_{fused} = \mathsf{Concet}(\mathsf{Z}_g, \mathsf{Z}_{v})$. Once we have the resulting concatenated output, we then forward the output to a multi-layer perceptron network, which is basically made of Rectified Linear Unit (ReLU) activation function and layer normalization procedures at each layer. We can describe the overall process as:
\begin{equation}
    \begin{array}{l}
    \mathsf{MLP}_1 = \mathsf{ReLU}(\mathsf{LayerNorm}(\mathsf{Z}_{fused}\mathbf{W}_1 + \mathsf{b}_1)) \\
    \mathsf{MLP}_2 = \mathsf{ReLU}(\mathsf{LayerNorm}(\mathsf{MLP}_1\mathbf{W}_2 + \mathsf{b}_2)) \\
    \hat{\mathsf{b}} = \mathsf{Softmax}(\mathsf{MLP}_2\mathbf{W}_3 + \mathsf{b}_3),
    \end{array}
\end{equation}
    where, $\mathbf{W}$ and $\mathsf{b}$ are the corresponding weights and biases, which will be learned during training. At the final layer, the $\mathsf{Softmax}(\cdot)$ function is applied to make a normalized output into a probabilistic prediction, which can be essentially interpreted to determine the $k$ best beams.
    
\section{Experiments and Performance Assessment}
    In this section, we present the experimental setup in details, subsequently, we also present the performance results.

\subsection{Experimental Setup and Network Training}
    For training and performance evaluation, we consider the DeepSense 6G dataset \cite{morais2025deepsense}, which is a collection of coexisted multi-modal sensor observations captured from real-world wireless environments. In consistent with our considered system model, we particularly adopt the vehicle-to-vehicle (V2V) communication scenarios 36, 37, 38, and 39 from this dataset. For each V2V scenarios, the testbed setups with transmitter and receiver units operating at 60 GHz bands are implemented to collect the data at different locations, such as Tempe, Phoenix, and Scottsdale of Arizona. In particular, the mmWave phased arrays at receiver units employ $16$-element ($N_{rx} = 16$) phased array antennas with an over-sampled $64$ beamsteering codebooks ($|\mathbfcal{C}_{rx}| = 64$), while the transmitter units have one antenna element ($N_{tx} = 1$) to make the omnidirectional transmission towards the receivers.
    
    In this work, we conduct a series of experiments by implementing the designed multi-modal model with the considered dataset on an Intel Core i7-10875H CPU and NVIDIA GeForce RTX 2080 Super GPU based computing platform. Specifically, we use Pytorch 1.13.1 framework in Python 3.7 and CUDA toolkit 11.7 to develop the model on this platform. During training of proposed multimodal fusion model, we utilize the adaptive moment estimation (Adam) optimizer with a learning rate of $0.001$ and weight decay of $1 \times 10^{-4}$ for faster convergence while minimizing the loss function. Besides, we randomly partition the data resources of each scenario into training, validation, and testing sets using a ratio of 6:2:2. Following these, we then train our model over $20$ epochs with a batch size of $16$ for each to get the decent accuracy. 

\subsection{Performance Assessment}
    Upon completing the training processes, the following two key matrices are utilized to assess the performance of the proposed model.
    
    \textit{1) Top-$k$ Accuracy:} is defined as the number of correct top-$k$ most likely outcomes with respect to true beams by the model. This metric essentially presents the the model's ability on making accurate predictions. Defining the top-$k$ accuracy as $Acc_{top-k}$, it can be given by:
\begin{equation}
    Acc_{top-k} = \frac{1}{N_{test}}\sum_{n=1}^{N_{test}} \left[\mathds{1}(\hat{\mathsf{b}}_n = Y^{gt}_n)\right],
\end{equation}
    where, $\mathds{1}(\cdot)$, $N_{test}$, $\hat{\mathsf{b}}_n$ and $Y^{gt}_n$ are the indicator function, total number of test samples, predicted beams at $n$-th test sample, ground-truth beams at $n$-th test sample, respectively.
    
    \textit{2) Average Power Loss:} is calculated from the average ratio between the downlink received powers with top-$k$ predicted beams and the maximum received powers possibly achieved at the link while considering the noises in realistic scenarios. In decibels (dB) scale, it can be expressed as:    
\begin{equation}
    APL[\text{dB}] = - 10 \log_{10} \left( \frac{1}{N_{test}}\sum_{n=1}^{N_{test}}\frac{\hat{\mathsf{p}}_n-\mathsf{p}_o}{\mathsf{p}_n^{gt}-\mathsf{p}_o}\right),
\end{equation}
    where, $\hat{\mathsf{p}_t}$, $\mathsf{p}_t^{gt}$, $\mathsf{p}_o$ are the corresponding powers from predicted beams at $n$-th test sample, the corresponding powers ground-truth beams at $n$-th test sample, and noise powers, respectively.
    
    With these performance matrices, we then perform a comparative study by evaluating the performance assessment with two established uni-modal data based approaches, hence no fusion, as baselines. In the first baseline, we consider the work in \cite{morais2023position}, which utilizes only a single GPS sensing coordinates modality. Further, for the second baseline, we inspire from the idea presented in \cite{kim2024computer}, where extracted features on vision modality has been used to do the same task. These baselines help us to observe how multi-modality fusion can improve prediction accuracy.
    
    In Fig. \ref{fig: Performances}, we illustrate the comparison among our proposed approach and the baselines across all considered scenarios in terms of accuracies and average power losses. Specifically, we can observe from the results in \ref{fig: Performances}(a)-(d) for the model accuracies that the baseline 1 and 2 on individually (that is, without fusing with each other) performs the worst, essentiality leading to many wrong beam directions. Notably, after performing the multi-modal fusion, the performance gets better than baselines as expected. For example, the V2V-day scenario achieves the maximum beam prediction accuracy of $77.58$\% with the proposed approach while predicting the top $15$-beams, whereas the baseline 1 and baseline 2 attain $52.68$\% and $73.74$\%, respectively. Consequently, we can further observe in \ref{fig: Performances}(e)-(h) that the proposed solution on all considered four scenarios experiences a significant reduction in terms of the average power losses with respect to the baseline 1 and baseline 2. For instance, the proposed approach incurs only $3.16$~dB average power loss in V2V-day first scenario with top $15$ beams, compared to $5.32$~dB and $3.80$~dB for baseline~1 and baseline~2, respectively. Besides, the average per sample inference execution time ranges between $0.5$ ms and $1.0$ ms.

\section{Conclusion}
    In this paper, we have presented a beam selection solution for $60$ GHz mmWave enabled connected vehicles. Based on the out-of-band contextual sensing information observed from the surroundings along with position information, we have designed a transformer based multi-modal fusion framework to predict top-$k$ beams, that is a subset of beams. Subsequently, we have evaluated the proposed framework with practically collected wireless datasets in diverse scenarios. In particular, our findings have highlighted the competitive performances of our proposed framework in terms of prediction accuracies and average power losses, marking the ability to meet high throughput and low latency demands of connected vehicles. Perhaps, exploring how to incorporate with emerging multimodal foundation models could be considered as a future work.
    
    \textbf{Acknowledgment:} This research is partly supported by National Science Foundation (NSF) under the grant numbers \# 2010366 and \#~2140729.

\ifCLASSOPTIONcaptionsoff
  \newpage
\fi

\bibliographystyle{IEEEtran}
\bibliography{bibliography.bib}

\end{document}